# SCALE-FREE FORAGING BY PRIMATES EMERGES FROM THEIR INTERACTION WITH A COMPLEX ENVIRONMENT

Denis Boyer[1], Gabriel Ramos-Fernández[2*], Octavio Miramontes[1], José L. Mateos[1], Germinal Cocho[1], Hernán Larralde[3], Humberto Ramos[1], Fernando Rojas[4]

[1] Departamento de Sistemas Complejos, Instituto de Física, Universidad Nacional Autónoma de México, Apartado Postal 20-364, 01000 México DF, México.

[2] Centro Interdisciplinario de Investigación para el Desarrollo Integral Regional, Unidad Oaxaca, Instituto Politécnico Nacional, Calle Hornos 1003, Santa Cruz Xoxocotlán, 71230 Oaxaca, México.

[3] Centro de Ciencias Físicas, Universidad Nacional Autónoma de México, Cuernavaca, 62210 Morelos, México.

[4] Facultad de Ciencias Físico Matemáticas, Universidad Autónoma de Puebla, Av. San Claudio y Río Verde, Col. San Manuel, 72570 Puebla, Puebla, México.

*Author for correspondence (ramosfer@sas.upenn.edu)

**Scale-free foraging patterns are widespread among animals. These may be the outcome of an optimal searching strategy to find scarce, randomly distributed resources, but a less explored alternative is that this behaviour may result from the interaction of foraging animals with a particular distribution of resources. We introduce a simple foraging model where individual primates follow mental maps and choose their displacements according to a maximum efficiency criterion, in a spatially disordered environment containing many trees with a heterogeneous size distribution. We show that a particular tree size frequency distribution induces non-Gaussian movement patterns with multiple spatial scales (Lévy walks). These results are consistent with field observations of tree size variation and spider monkey (*Ateles geoffroyi*) foraging patterns. We discuss the consequences that our results may have for the patterns of seed dispersal by foraging primates.**

## 1. INTRODUCTION

Many animal species, including albatrosses (*Diomedea exulans*; Viswanathan *et al.* 1996), jackals (*Canis adustus*; Atkinson *et al.* 2002) and spider monkeys (*Ateles geoffroyi*) (Ramos-Fernández *et al.* 2004), move in their environment along apparently erratic trajectories which can be accurately described as Lévy walks. These are random walks composed of a sum of independent steps (or sojourns), but with markedly non-Gaussian statistics due to a diverging mean square of the step length. That is, the probability distribution $P(l)$ for the length $l$ of each sojourn is broadly distributed and self-similar over a wide range of scales. Normal (Brownian) random walks are characterized by steps with finite mean square length which contribute by roughly the same amount to the overall displacement, giving rise to Gaussian diffusion. In contrast, Lévy walks generate anomalous (namely, faster) diffusion as they are dominated by very large, though infrequent, steps (Bouchaud & Georges 1990; Shlesinger *et al.* 1993; Klafter *et al.* 1996). Recently, we have determined the distribution of the sojourns given by fruit-eating spider monkeys foraging in a semi evergreen tropical forest in Yucatan, Mexico (Ramos-Fernández *et al.* 2004). The results show power-law behaviour $P(l) \sim l^{-\alpha}$, with an exponent $\alpha \approx 2.2$, which suggests that these foraging movements may indeed be described by Lévy walks (for Lévy processes, $1 < \alpha \leq 3$). Similar results, with $\alpha$ also close to 2 have been reported for other animals in other contexts (Viswanathan *et al.* 1999).

Lévy foraging behaviour may be the outcome of an optimal searching strategy to find scarce, randomly distributed resources (Viswanathan *et al.* 1999). A less explored alternative is that this behaviour may be the outcome of the distribution of resources themselves. It is well known that many animals (bees, rodents, primates) do not forage randomly but rely instead on cognitive maps to navigate their environment (Collett *et al.*

1986; Garber 1989; Dyer 1994). These maps may contain information on the location of different targets and the geometric relationships between them (Kamil & Jones 1997). Animals are also able to evaluate the amount of food present at different locations within their environment (Shettleworth *et al.* 1988; Janson 1998;) and in some cases integrate this information with their spatial knowledge (Janson 1998). In the field, we have observed that spider monkeys follow regular routes to travel between feeding sites within a limited area of approximately 2 km$^2$ (Ramos-Fernández & Ayala-Orozco 2003; Valero 2004).

Here we introduce a novel foraging model, in which the territory is composed of many targets (food patches) with varying sizes and in which the spatial structure can be varied; foragers know the location and size of the targets and adhere to a simple foraging strategy (i.e. maximizing food intake in a minimum travelled distance). We use the model to explore the conditions that lead to Lévy foraging patterns. Our numerical simulations show that a particular target size frequency distribution, similar to the tree size frequency distribution measured in the field, induces the most non-Gaussian trajectories. The corresponding step length distribution is a Lévy law with $\alpha = 2$, in agreement with the collected foraging data (Ramos-Fernández *et al.* 2004). The results are also consistent with the observed distribution of waiting times, i.e. the time spent feeding in trees between two moves (Ramos-Fernández *et al.* 2004). We discuss the relevance of our findings for understanding seed dispersal patterns by frugivorous primates.

## 2. THE BASE MODEL

We model the foraging environment as a two-dimensional square domain of area set to unity for convenience, containing $N$ point-like targets ($N >> 1$). In a first approximation, targets are randomly and independently distributed in space (Poisson process). The size of the system can be thought of as the size of the territory of a group of monkeys. Targets represent the trees with fruits that monkeys eat; we assign to each target $i$ a random integer $k_i \geq 1$ representing the target's size, or fruit content. Recent work (Enquist & Niklas 2001, Niklas *et al.* 2003) has found that in many tropical and temperate forests, the probability $p(k)$ of observing a tree of size $k$ (estimated as the diameter at breast height, see more discussion on this in the Result Section) falls as an inverse power law. Thus, we assume that $k_i$ is distributed according to a inverse power-law probability distribution

$$p(k) = Ck^{-\beta}, \quad C = 1/\sum_{k=1}^{\infty} k^{-\beta} \tag{1}$$

where $C$ is the normalization factor, and $1 < \beta < \infty$ is a fixed exponent characterizing the environment, being the only parameter of the model. If the exponent $\beta$ is close to 1, $p(k)$ decays slowly with $k$, implying that the range of target size is very broad, with essentially all sizes present. In contrast, when $\beta \gg 1$, practically all targets have the same size $k_i = 1$ and the probability to find larger ones ($k_i = 2, 3 \ldots$) is negligible. Many types of forests have been found to be characterized by values in the range $1.5 \leq \beta \leq 4$ (Enquist & Niklas 2001, Niklas *et al.* 2003).

We now consider a forager located at a starting point near the centre of the domain. The forager knows the location and size of all targets in the system. The following rules of motion are recursively implemented:

   (a)    the forager located at the target number $i$ will move in straight line to a target $j$ such that the quantity $l_{ij} / k_j$ is minimal among all available targets $j \neq i$ in the system, where $l_{ij}$ is the distance separating the two targets and $k_j$ is the size of target $j$;

(b)      the forager does not choose an already visited target, as it is assumed that visited targets no longer contain fruits.

According to rule (a), valuable targets (large $k$) may be chosen even if they are not the nearest to the monkey's position. The quantity $l/k$ roughly represents a cost/gain ratio for a move. Our assumption that foragers know the position and size of all targets could be relaxed by assuming that they only know a random subset of the total. Those known targets would still vary in size according to the same overall distribution (equation 1), and the results reported here would not change. We emphasise that, once the random environment is set and an initial position chosen, the trajectory, although complicated, is not random but deterministic. In the Appendix, we develop two modifications of our model designed to test the robustness of the results. Further improvements, not considered here, could relax rule (b) and allow revisiting after a period of time that is sufficient for a patch to replenish.

RESULTS OF THE BASE MODEL

We first analyse the average properties of the trajectories composed of many steps, and their variations with the type of environment, namely, the resource exponent $\beta$. The numerical results can be summarized from Figure 1a.

(i)      For $1 \leq \beta \leq 2$, the sojourn length frequency distribution, denoted as $P_\beta(l)$, is very broad and is only limited by a characteristic length scale of order of the system size ($L=1$), which is much larger than the mean separation distance $l_0$ between two nearby targets ($l_0 \sim N^{-1/2}$).

(ii)      For $2 \leq \beta \leq 4$, $P_\beta(l)$ is broad but does not significantly depend on the size of the system: in the sub-range $3 \leq \beta \leq 4$, it is very well fitted by a power-law distribution for large $l$, $P_\beta(l) \sim l^{-a}$, with a measured exponent $\alpha \approx \beta - 1$, while important deviations from power-law behaviour appear in the sub-range $2 \leq \beta < 3$.

*(iii)*      For $\beta > 4$, $P_\beta(l)$ decays faster than $l^{-3}$, meaning that sojourns between nearby targets entirely dominate the statistics.

These results can be understood qualitatively as follows. At large values of the resource exponent $\beta$, all targets are practically identical in size and the forager travels to the nearest unvisited one (Lima *et al.* 2001), performing trajectories resembling those already observed in some herbivores (Gross *et al.* 1995). In contrast, for small $\beta$, the largest targets available in the system are much larger than average and also relatively numerous. As the distance from any point of origin increases, there is a higher probability of finding increasingly larger targets, so that sometimes the forager will decide to visit them in spite of the distance it has to travel. In that lower exponent range, the length of a long sojourn can be of order of the system size, with the resulting trajectories resembling those of a random walker trapped in a finite domain. Lévy-like trajectories occur at intermediate values of $\beta$, for which large targets are indeed present, but are still scarce and far apart from each other. In this regime, the trajectories are not strongly affected by the finite size of the system except after very long times. In Fig. 2a, we show numerical results for the relative step length fluctuations as a function of $\beta$, along with typical trajectories. The resource size distribution with $\beta_c = 3$ is special because it induces the largest relative fluctuations on the length of the sojourn $l$. These trajectories contain the largest number of length scales, characterized by $P(l) \sim l^{-\alpha}$, with $\alpha \approx 2$ (see Fig. 1a).

It is worth noting that the special state at $\beta_c = 3$ most resembles the trajectories described by spider monkeys in the field (Ramos-Fernández *et al.* 2004). Additionally, this resource exponent value is close to that which characterizes the real variation in tree size, as measured in the field in an area close to the study site. The tree-size distribution at this site can be fitted as a power-law with exponent value $2.6 \pm 0.2$ (Figure 2b). These results are based on measurements of the diameter at breast height (DBH), which is commonly regarded as one of the most accurate methods for estimating fruit abundance of tropical tree species (Chapman *et al.* 1992). It is observed in some examples that the mass of reproductive structures is roughly a linear function of DBH (McDiarmid *et al.* 1977; Snook *et al.* 2005), as well as of tree size (Crawley 1997; Fenner & Thompson 2005).

Our model accounts for another feature that illustrates the interaction between spider monkeys and their environment. In the field, once a monkey has stopped at a given tree, it tends to stay there for a random time interval, $\tau$, that is also distributed as a power-law (Ramos-Fernández *et al.* 2004). In spite of the fact that most waiting times are short (5 to 10 min.), a significant amount of them are greater than two hours. As reported in Figure 3a, the measured waiting time distribution of spider monkeys can be fitted by the law $\psi(\tau) \sim \tau^{-w}$ with $w \approx 2.0$ (Ramos-Fernández *et al.* 2004). We then assume that the time spent on a tree is proportional to the amount of food available at that tree, which at the same time is proportional to the tree size. In Figure 3a, we also show the frequency distribution of the size of the targets visited by the model forager, noted as $P_{\beta=3}^{(v)}(k)$, for the special value $\beta_c = 3$ of the model. Such distribution is not straightforward: the walker visits a subset of targets that have relatively larger sizes compared with the overall distribution $p(k)$ of equation (1). This is due to the attractiveness of larger targets in the choice process. We observe $P_{\beta=3}^{(v)}(k) \sim k^{-\gamma}$, with $\gamma \approx 2.0$. This numerical value is in quantitative agreement with the waiting time exponent $w \approx 2.0$ measured in the field, as would be expected under the assumption that the time spent feeding on a tree is proportional to its size or its fruit content.

Fruit-eating monkeys swallow the seeds of many tree species and deposit them away from the parent tree after an average transit time of 4.4 hours in the case of spider monkeys (Lambert 1998). During this time, a spider monkey may travel distances of tens to hundreds of meters. Indeed, in previous work (Ramos-Fernández *et al.* 2004) we observed that during the early morning period, when monkeys forage away from their sleeping trees, their mean displacement grows algebraically with time, as $t^{0.85}$. This growth rate is faster than ordinary diffusion, but slower than straight line motion. In the model, one can similarly define the mean displacement of the walker as the average quantity $< |\boldsymbol{R}(t_0 + t) - \boldsymbol{R}(t_0)| >$, where $t$ is the total duration of the walk, $\boldsymbol{R}$ is the walker's position vector and $t_0$ an arbitrary origin time. A walker arriving at a target of size $k$ stays there for a time equal to $k$ (see above and Figure 3a) before moving toward the next target at a constant speed, e.g., $l_0$ per unit time. In Figure 3b, we have plotted the resulting displacement as a function of time $t$ for various resource exponents $\beta$. We observe that the mean displacement of the forager away from its starting point after any given time is maximal when the environment has the target size frequency distribution exponent $\beta = 3$. This property is independent of the forager's speed: we have verified that, for any fixed value of the speed of motion between two targets, the displacement is always maximal when $\beta = 3$. This result still holds if the distribution (1) is truncated at a particular size $k_{max}$ (if $p(k > 100) = 0$, for instance).

Qualitatively, this result implies that at low $\beta$ values ($< 3$) the large quantity of large targets slows down the forager's progression through the system. At large $\beta$ values ($>3$), on the other hand, most targets are small and very large ones are so scarce that they have little impact on the trajectories. In this case, waiting times are short, but so are the length of the steps. An intermediate situation (not so long waiting times, long step lengths) is achieved for $\beta = 3$, the value at which the mean displacement, and therefore dispersal, is maximal.

DISCUSSION

Despite its simplicity, the model shows a rich variety of behaviour. By varying its main parameter, which describes the decay of the tree-size frequency distribution, the trajectories of a forager following a simple optimization rule can differ widely. The agreement found between the field exponents (for step length, tree size and waiting time distributions) and their theoretical values at the special parameter $\beta_c = 3$ suggests that the model correctly captures the interactions between spider monkeys and their environment. At these values, forager trajectories contain the largest possible fluctuations regarding the length of constituent sojourns and thus can be correctly characterized as Lévy walks.

Viswanathan et al. (1999) suggested that Lévy walks may be part of an optimal searching strategy to find scarce, randomly distributed resources. Their argument is based on the fact that, compared to Brownian foragers (who perform walks with a constant length of constituent sojourns), Lévy foragers would reach new, unvisited areas in shorter times as well as having a smaller probability of reaching areas already visited. However, many animal species, from insects to primates, have been shown to possess a sophisticated knowledge of resource location (e.g. Garber 1989; Dyer 1994). Our results suggest that Lévy walks arise as a consequence of food intake maximization in a spatially disordered, heterogeneous environment where the location of resources is at least partially known.

A crucial assumption of our model is that spider monkeys rely on mental maps in their foraging, as they move from one fruiting tree to another. Recent work (Valero 2004), carried out in the same study site as the one where the movement patterns were first studied (Ramos-Fernández *et al.* 2004), found that spider monkeys can orient their straight-line movements toward existing fruiting trees at distances of up to 1500m. These distances appear too great to have been the result of detection by sight alone and instead suggest that spider monkeys use some kind of mental representation of the location of current feeding sources (Valero 2004). A variant of our model (see Appendix) introduces a degree of error of 65% in the foraging rule employed by foragers, effectively eliminating, in many cases, the best option available. Still, the statistics of the foraging patterns remain unaffected.

The distribution of resources on which spider monkeys actually rely on is close to an inverse power law with exponent $\beta = 3$. Similarly, tree-size distribution measurements reported elsewhere (Enquist & Niklas 2001, Niklas *et al.* 2003) in several forest types, have shown that typical exponent values are in the range $1.5 \le \beta \le 4$. Our results highlight some of the consequences that power-law size distributions in tree communities can have on the foraging patterns of animals that utilise them as a resource.

The phenology of trees deserves special attention, as the diet of foragers largely depends on the timing of fruiting (Rathcke & Lacey 1985). So far, we have not included different species in our model, and it is clear that there are times when more abundant or larger tree species would provide larger amounts of fruit (Newstrom et al. 1994). This could imply that the total number of feeding trees, as well as their size frequency distribution (*e.g.* the exponent $\beta$), at any one time might not remain constant across seasons. The model can provide testable predictions on the way in which foraging trajectories would vary. For instance, the results above still hold if fruiting trees are modelled as a random subset of the overall set of targets. Additionally, the spatial structure of feeding trees could change with time, a feature that can be incorporated into the model with the help of the Markov point processes mentioned in the Appendix.

Our results may have important consequences for the ecology of the trees that spider monkeys use, particularly on the seed dispersal patterns. Long-distance seed dispersal has been identified as a key determinant of the spatial structure and species composition of tree communities (Janzen 1970; Nathan & Muller-Landau 2000; Chave *et al*. 2002). Frugivorous primate species, in particular, are important seed dispersers as they have been shown to disperse the seeds of twice as many species as birds (Clark & Poulsen 2001), affecting the spatial patterns of seed abundance at various spatial scales (Julliot 1997; Wehncke *et al.* 2003). These primate species are known to disperse seeds at modal distances of 200-500 m, although the longest observed dispersion events can go up to 840 m for capuchin monkeys (Wehncke *et al.* 2003) or 1 km for woolly monkeys (Stevenson *et al.* 2000). Such "fat tail" in the probability of seed dispersion as a function of distance from the source has been suggested by Clark *et al.* (1999) to be a more appropriate representation of the dispersion probability function.  In our model, we find that the movement patterns described by foragers could disperse seeds in precisely this way. By moving in a series of short sojourns followed by rare but very long ones, seeds would be deposited frequently near the source but also, infrequently, at very long distances from it.

The results presented in the Appendix, on the other hand, show that dispersal is also sensitive to the existing spatial structure of trees at short-to-intermediate length scales. Under the same foraging rules and size distribution of resources, seeds are likely to be spread further in a community that is spatially random/uniform (with the Poisson forest as a limiting case), than in a spatially structured/heterogeneous community (*e.g.* clumped). In order to tell whether foragers would be contributing to the evolution of the tree population toward a particular size or spatial structure, a description of forest dynamics should be further coupled to the present model. For instance, many factors would influence the growth success of a seed that has arrived at a given site: the distance from parent tree (Janzen 1970; Harms *et al.* 2000),  nutrient availability, competition with other seeds or predation (Jordano 1992; Nathan & Muller-Landau 2000). Our model does not specifically address the issue of species diversity. However, it gives some insights on a mechanism (among many) which can have an impact on diversity at intermediate scales in communities composed of individuals with known size and spatial distributions.

In summary, we have identified a novel mechanism by which a realistic, scale-invariant distribution of tree size generates Lévy-walk foraging movements as an emergent pattern. These findings have been supported by various field measurements. Our results should be considered as an alternative explanation of the prevalence of Lévy walks in animal behaviour, especially of species with sophisticated knowledge about their environment. On the other hand, these foraging patterns may have important consequences for the dynamics of tree communities.

## ACKNOWLEDGEMENTS

We are indebted to Robert M. May, Annette Ostling, Tim Coulson, Michael P. Hassell, François Leyvraz, Eliane Ceccon, Gregory S. Gilbert, Barbara Ayala, Michael F. Shlesinger and Joseph Klafter for fruitful discussions, and to the late Ingrid Olmsted from the Centro de Investigación Científica de Yucatán for the data on tree size in a tropical forest in the Yucatan peninsula. This work was supported by CONACYT grants 40867-F

and G32723-E, SEMARNAT-CONACYT grant 0536 and DGAPA grants IN-111000 and IN-100803.

APPENDIX: Robustness of the model

The model above relies on a few basic assumptions. To further investigate the robustness and the generality of its predictions, we also present two variants including more specific hypotheses.

*Variant I: Tree correlations.* Trees in tropical forests are usually not randomly distributed in space. They tend to be aggregated in many cases and clumping is particularly strong among conspecifics (Crawley 1997; Condit *et al.* 2000). Spatial correlations still exist across different species, although they are much lower (Pélissier 1998). The diet of spider monkeys is composed of hundreds of different species (van Roosmalen & Klein 1987; Ramos-Fernández & Ayala-Orozco 2003). Even in a single month, spider monkeys may have access to more than 20 different species (Ramos-Fernández, unpublished data). Even though the random (Poisson) assumption made above may be not too far from reality as far as the overall set of available trees is concerned, it is instructive to further take into account an *a priori* existing stand structure to study its impact on forager trajectories. In this first variant of the model, the target size distribution in Eq. (1) and the foraging rules are unchanged, but the targets composing the system are now correlated in space.

Markov point processes are phenomenological methods that are widely used in forestry statistics to generate correlated spatial patterns (Ripley 1977; Stoyan & Penttinen 2000). A pair wise interaction $\phi(l_{mn})$ is introduced, depending on the distance separating two trees $m$ and $n$, for instance, $\phi$ being related to the likelihood that two trees are found at a given distance. Further, the forest is modelled by $N$ points ($N=7.2 \ 10^4$ here) such that the probability density of finding them at positions $\{x_1, ..., x_N\}$ is set as proportional to exp[-$\Sigma_{m<n} \phi(l_{mn})$], the sum running on all possible pairs of points. Representative patterns are obtained after many computer iterations by using a depletion-replacement algorithm usual in Monte-Carlo calculations (Binder & Heermann 2002). We choose a standard shape for $\phi(l)$: the pair interaction is supposed to be infinitely repulsive (hard-core) at short distances, attractive at intermediate distances, whereas targets do not interact directly at larger distances. More specifically, $\phi(l) = \infty$ for $0 < l < \sigma$; $\phi(l) = -\gamma$ for $\sigma < l < R$; and $\phi(l) = 0$ for $l > R$, where $\sigma=0.5d$, $R=3d$, with $d=4.47 \ 10^{-3}$. The hard-core repulsion produces a fairly regular pattern at short length scales, while the attractive potential (interaction strength $\gamma \geq 0$) generates positive correlations further away (clumps). Correlations become negligible for sufficiently large separation distances.

We then consider two cases: In Case *(1)*, any pair of targets $\{m, n\}$ interacts through a unique potential, with $\gamma = 0.1$ as described above. In Case *(2),* a pair of targets interacts with the potential of Case *(1)* only if the ratio of their sizes $k_m \ / \ k_n$ is not larger than 2 (with the convention $k_m \geq k_n$); if not, the value $\gamma=0$ is set in their interaction.

In Case *(1)*, the interaction between targets does not depend on their sizes: the method produces point patterns with spatial correlations, the sizes of neighbouring targets remaining uncorrelated like in the base model. Case *(2)*, on the other hand, favours configurations where targets of similar sizes are more likely to be clumped, producing an effective repulsion between targets of very different sizes. Inhibition is commonly observed in evergreen tropical forests between adult and young trees, in part due to competition for light (Pélissier 1998).

In Case *(2)*, the overall target-target pair correlation function $g(l)$, related to the Ripley's *K* function (Stoyan & Penttinen 2000), decays slowly toward the value unity with the separation distance $l$, and indicates larger clumps than in Case *(1)*: $g(l = \sigma) \approx 1.84$ and $g(l = 5\sigma) \approx 0.99$ for Case *(1)*, whereas $g(l = \sigma) \approx 1.43$ and $g(l =5\sigma) \approx 1.14$ for Case *(2)* with a size distribution exponent $\beta = 3$. (A Poisson process gives $g(l) = 1$ for any $l$.) The values above are of the same order of magnitude than the ones measured in tropical forest stands when species are not distinguished (Pélissier 1998).

*Variant II: Imperfect foraging efficiency*. In order to relax the assumption of perfect knowledge by the foragers, the second variant of the base model modifies the foraging rule (a), while the targets remain Poisson distributed in space and with independent weights. In this variant, we assume that the forager is inexact in evaluating the distance to a given target, as well as its fruit content. The forager located at target $i$ will evaluate the cost of a move to target $j$ as $l^*_{ij}/k^*_j$, where $l^*_{ij}$ and $k^*_j$ are subjective distances and fruit contents deviating from the real ones: $l^*_{ij} = l_{ij}(1+e\,\eta_j)$ and $k^*_j = k_j(1+e\lambda_j)$, with $e$ a constant lower than 1 and $\{\eta_j, \lambda_j\}$ two random numbers uniformly distributed in the interval [-1,1]. Before each move, a pair of random numbers $\{\eta_j, \lambda_j\}$ is attributed independently to each unvisited target $j$, making its attractiveness over or under-estimated. The forager moves to the target of lowest $l^*_{ij}/k^*_j$ among the unvisited $j$'s. The results presented here have been obtained with $e = 0.30$, corresponding to an error of typical amplitude of 65%.

We now discuss the results given by the two variants of the model which address its robustness. Figure 4 displays the frequency distribution of the length of the sojourns, the tree size distribution exponent being fixed to the value $\beta = 3$. Spatial correlations between targets modify the shape of $P(l)$ at short distances. However, the scaling law $P(l) \sim l^{-2}$ at large $l$ found for uncorrelated targets is not modified with the introduction of spatial correlations without size affinity (variant *I*-Case *(1)*). The average mean displacement of the forager is also unchanged, as shown in the Inset of Figure 4.

Interestingly, deviations from the base behaviour arise when targets are correlated in size (variant *I*-Case *(2)*), the frequency of long sojourns being noticeably lower. This feature can be interpreted by the fact that large targets (the ones that can be reached after a long sojourn) tend to be clumped in this case, which reduces the distance between them compared to the Poisson situation. When reaching a clump of large targets, the forager may not need to give long sojourns for a while. As a consequence, the mean displacement (and therefore dispersion) is significantly reduced compared to the Poisson case, as shown in the Inset of Figure 4.

When the foraging rule is imperfect and with a typical error of 65 % (variant *II*), no qualitative differences in the distributions are found with respect to the perfect, deterministic rule (Fig. 4).

FIGURE LEGENDS

Figure 1: **(a)** Normalized step length distribution $P_\beta$ for various resource exponents $\beta$, obtained from simulations with $N = 10^6$ targets in a square domain, and averaged over 10 independent environment realizations (in each run, the number of visited sites is small compared with $N$). The length $l_0 = N^{-1/2}$ is the average distance between two nearby targets. The curves $\beta = 3$ and 4 are translated upward for clarity. Solid lines are inverse power laws $l^{-2}$ and $l^{-3}$. The filled squares correspond to the monkeys' foraging patterns collected in the field. **(b)** Spatial map of a trajectory performed by a spider monkey in the field (detail; see Ramos-Fernández *et al.* 2004 for details on the field study).

Figure 2: **(a)** Fluctuation ratio $< l^2 > / < l >^2$ (mean square length to square mean length) associated with the step length distributions $P_\beta(l)$ of Fig.2, as a function of the resource exponent $\beta$. The vertical lines are guides to the eye. Insets: spatial maps of typical trajectories for $\beta = 3$ and 5. **(b)** Tree size frequency distribution in semi-evergreen medium forest in La Pantera, in the southeastern Yucatan peninsula, Mexico. This is the same forest type with the same species composition as the spider monkey study site. Data conform to a power law with exponent $\beta \approx 2.6$ (± 0.2 standard error). Adjustment was performed using a least squares regression. Data consist of the diameter at breast height of a total of 250 trees ranging from 10 to 63.4 cm. See (Cairns *et al.* 2003) for more details on the study site and procedures. Data were kindly provided by the Centro de Investigación Científica de Yucatán.

Figure 3(**a**): Waiting time distribution. Once a monkey has stopped at a tree, it stays there for a time $\tau$ before moving to another site. The measured waiting time distribution of spider monkeys in the field is plotted here as filled squares, and is fitted by an inverse power law, $\psi(\tau) \sim \tau^{-w}$ with $w \approx 2.0$ (one time unit represents a 5min interval; see Ramos-Fernández *et al.* 2004). Also plotted – as open squares – is the distribution $P^{(v)}_{\beta=3}(k)$ of the size of the targets *visited* by the walker in the model, at the particular exponent value $\beta = 3$. We observe $P^{(v)}_{\beta=3}(k) \sim k^{-\gamma}$, with the same value $\gamma \approx 2.0$. One iteration of the model is equivalent to 5 minutes in the field observations of spider monkeys. (**b**): Mean displacement of the model walker, $< |\boldsymbol{R}(t_0 + t) - \boldsymbol{R}(t_0)| >$, as a function of the duration of the walk *t*. A walker arriving at a target of size *k* stays there for a time $t = k$ before moving toward the next target at a constant speed ($l_0$ per unit time).

Figure 4: Frequency distribution of the sojourn length given by the different variants of the model as described in the Appendix, for $\beta$=3. Inset: Corresponding mean displacement as a function of time (same legends).

Page headings: Tree size scaling and foraging Lévy walks

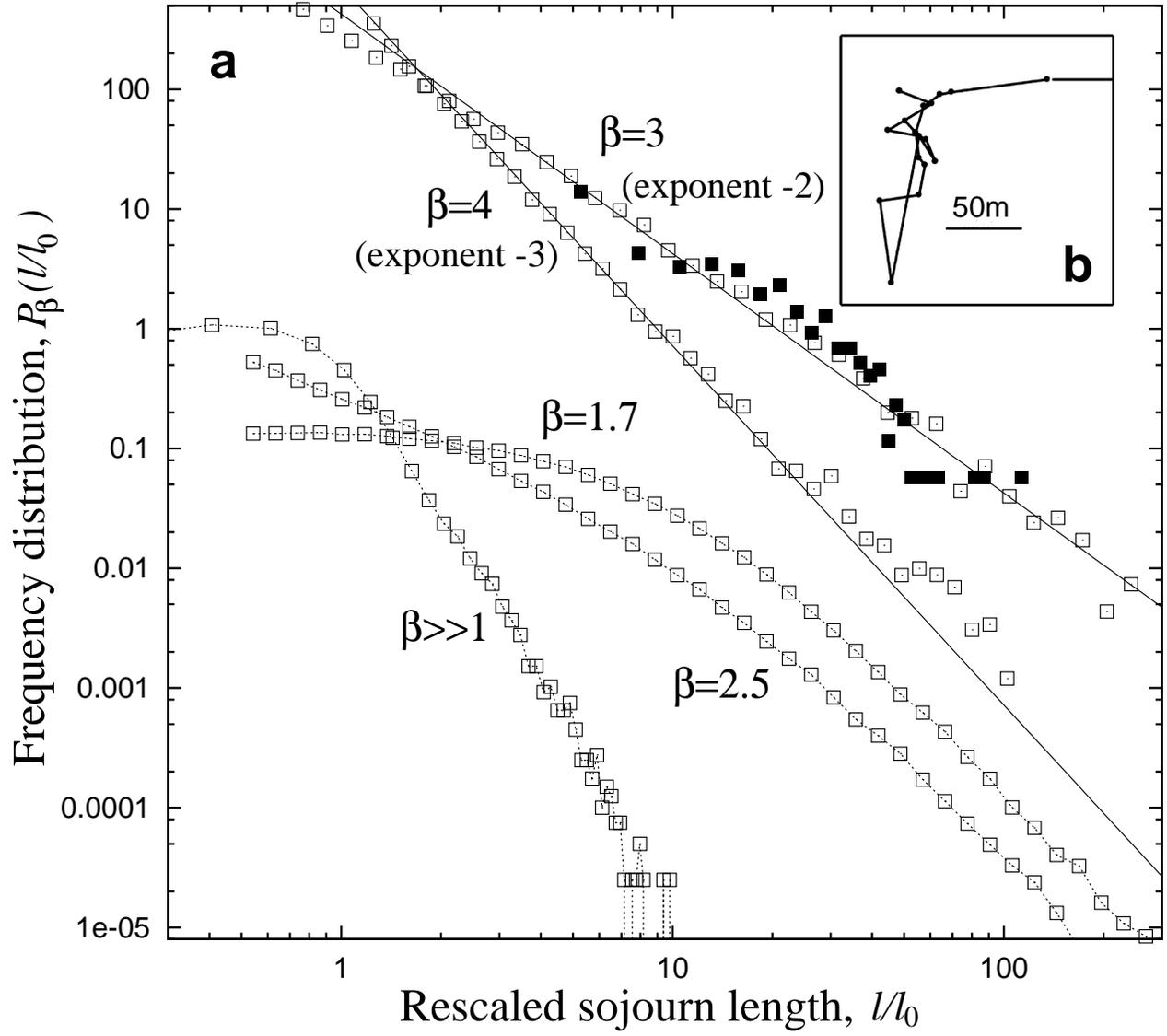

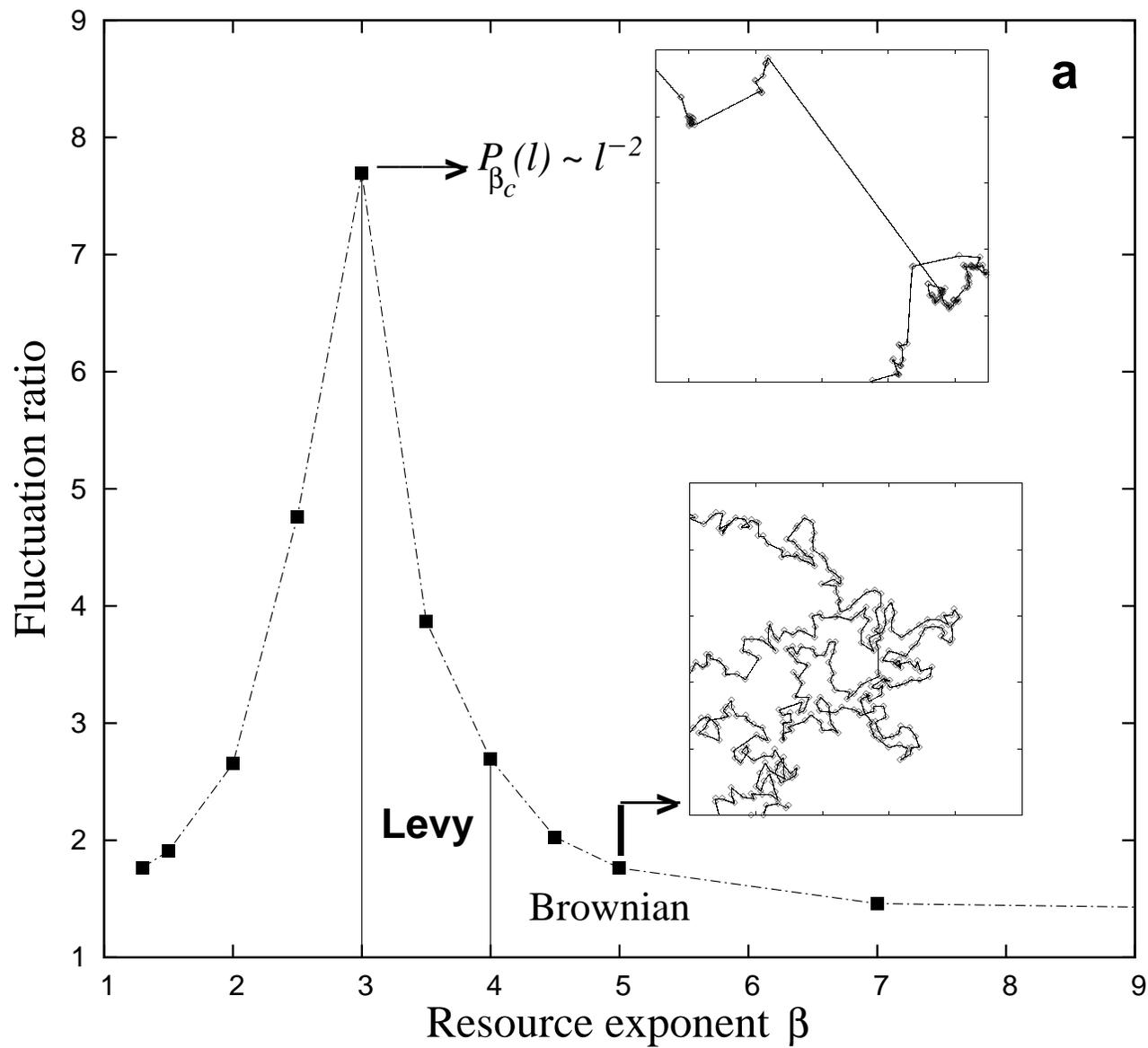

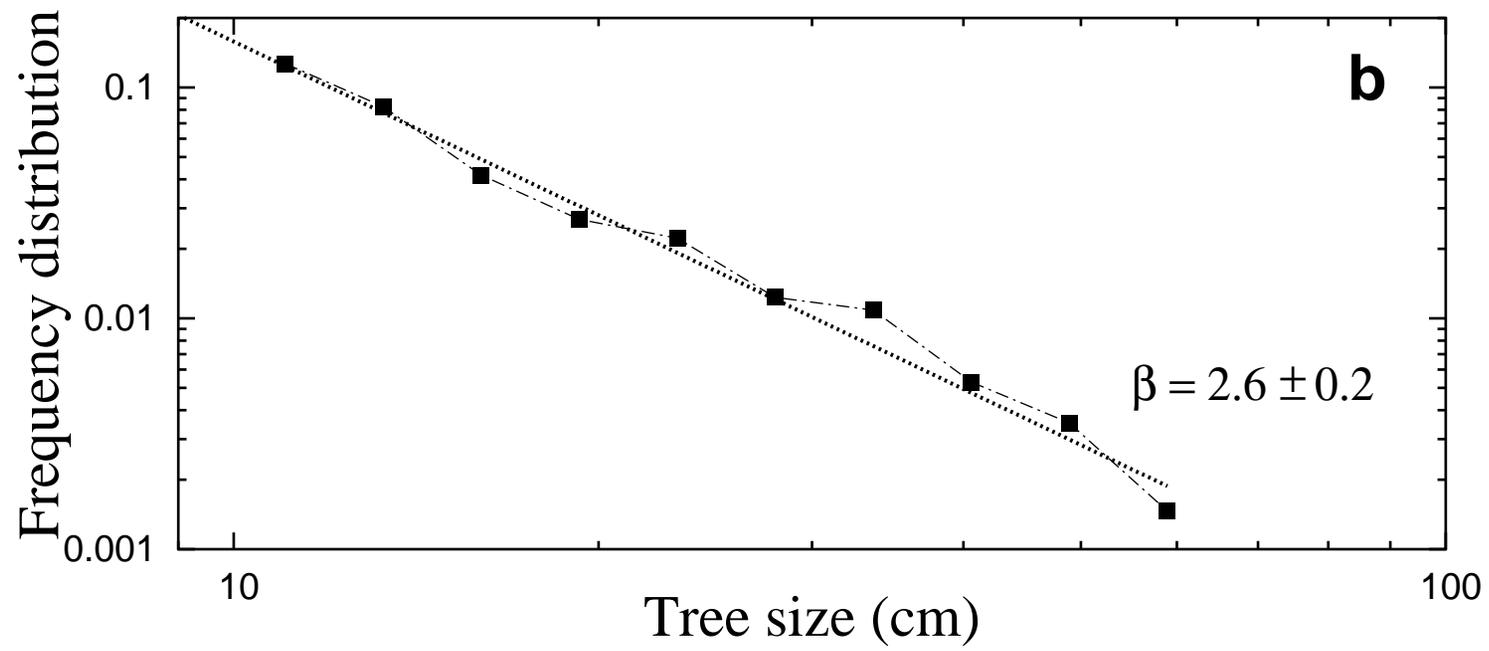

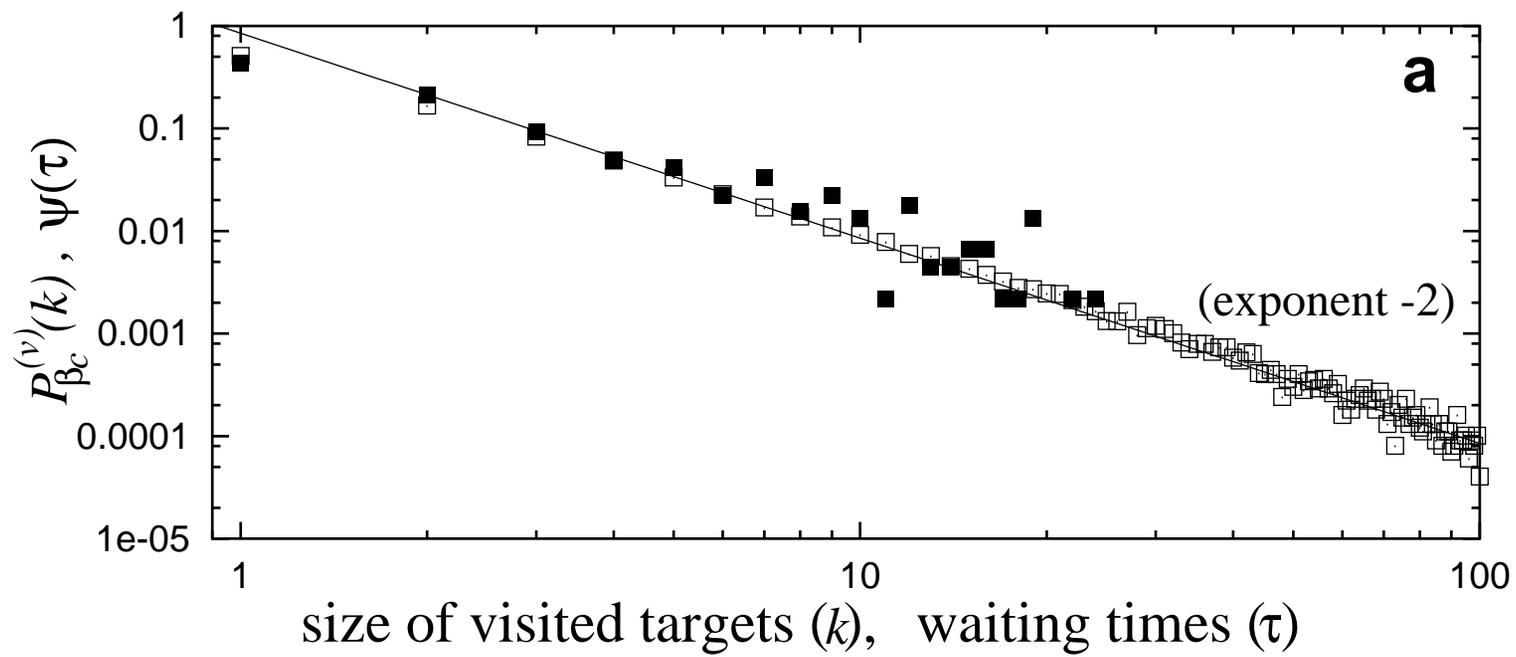

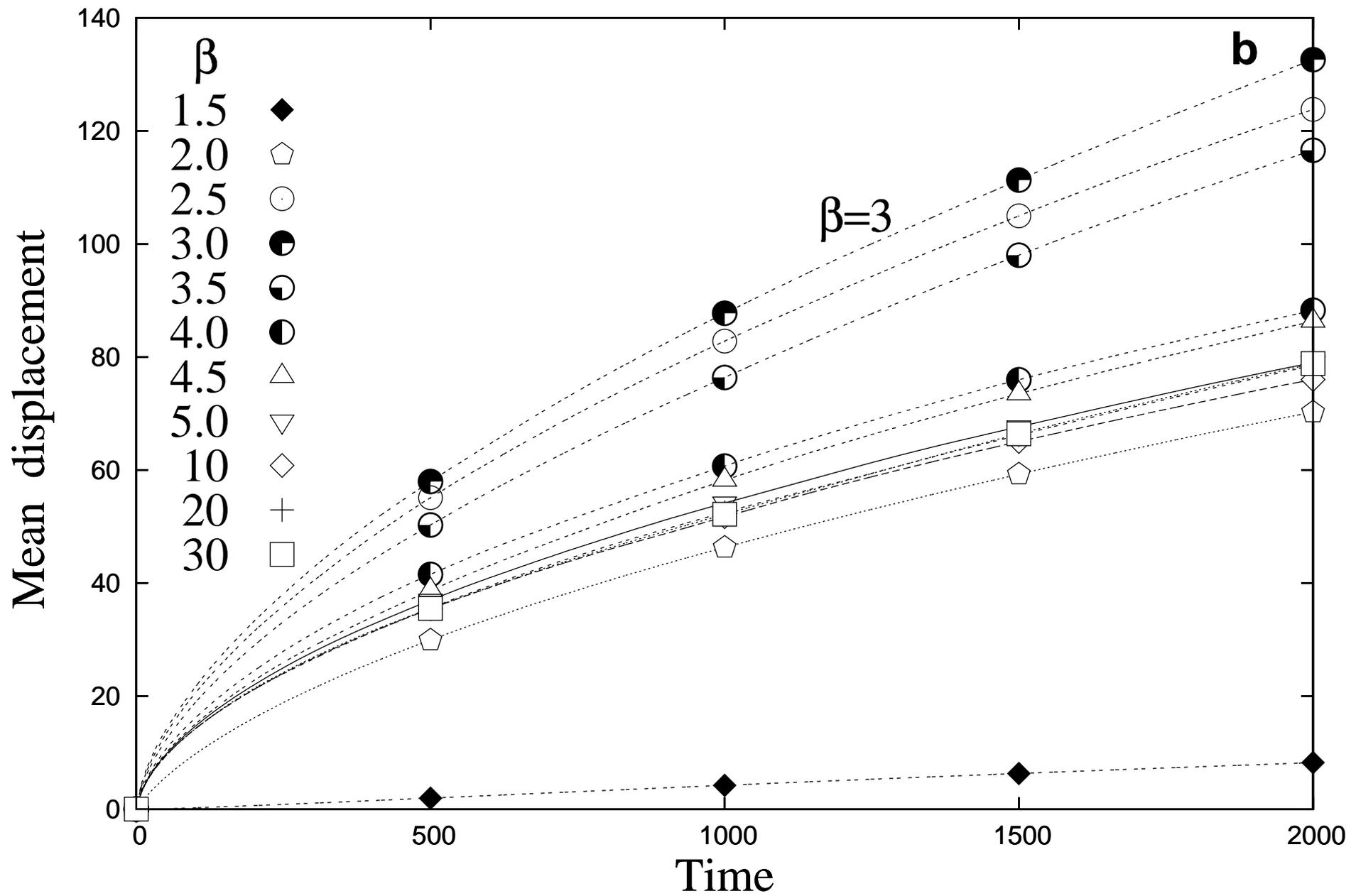

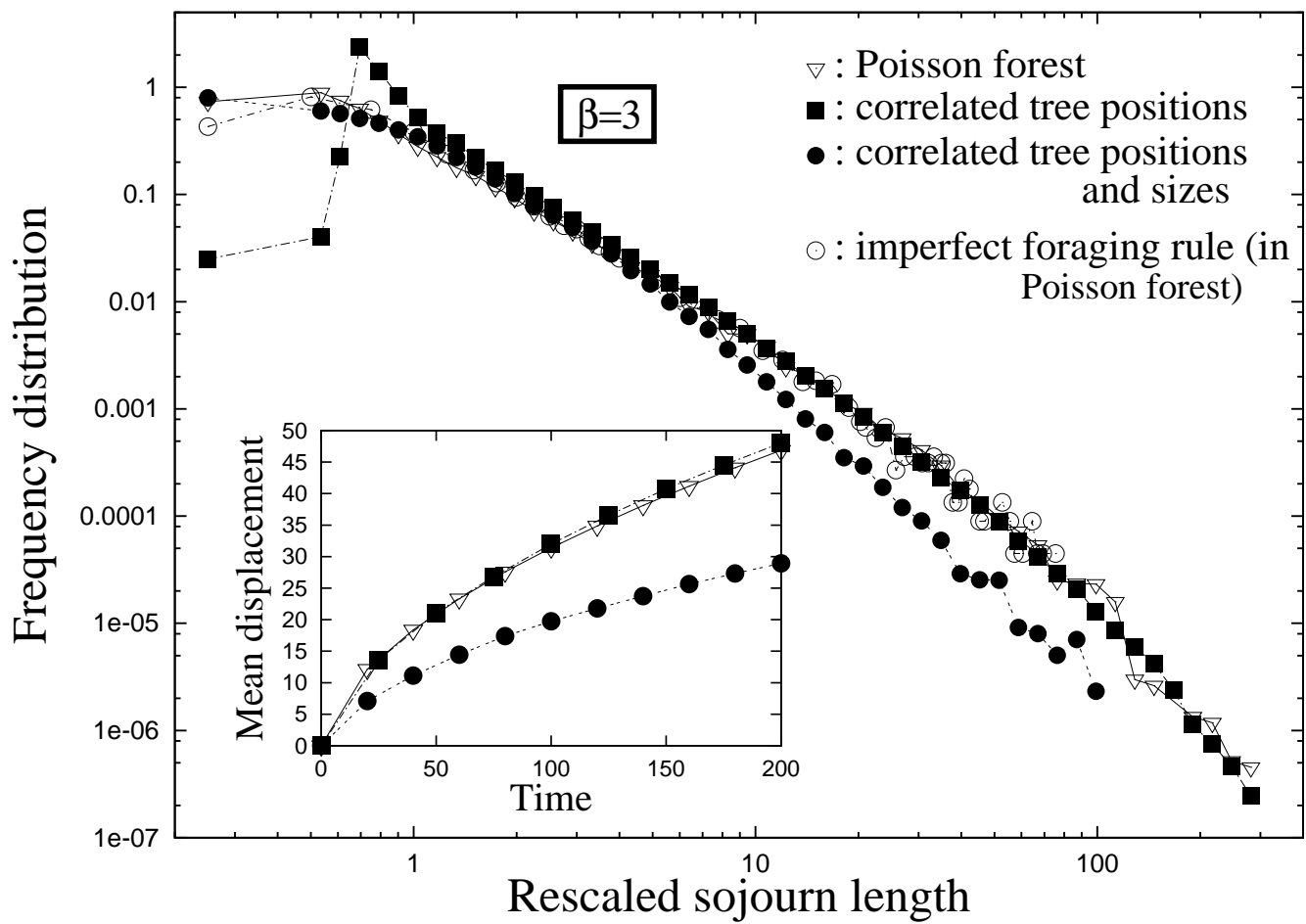